\documentclass[preprint,12pt]{elsarticle}

\usepackage{amssymb}
\usepackage{amsmath}

\usepackage[mathlines]{lineno}

\journal{Intermetallics}

\begin{document}

\begin{frontmatter}

\title{Oxygen-induced Fe surface segregation at the $L1_0$-FePd(001)/graphene heterointerface for spintronics devices: a first-principles study}

\author[kobe]{Mitsuharu Uemoto \corref{cor1}}
\author[kobe]{Naohiro Matsumoto}
\author[kobe]{Ryusuke Endo}
\author[ens]{Samuel Vergara}
\author[eeis, csrn]{Masaki Kobayashi}
\author[aist]{Hikari Shinya}
\author[imass,ias,toyama]{Hiroshi Naganuma}
\author[kobe]{Tomoya Ono}
\ead{uemoto@eedept.kobe-u.ac.jp}

\affiliation[kobe]{
  organization={ Department of Electrical and Electronic Engineering, Graduate School of Engineering, Kobe University},
  addressline={ 1-1 Rokkodai-cho, Nada-ku},
  city={Kobe},
  postcode={651-8501},
  country={Japan}
}

\affiliation[ens]{
  organization={ ENS Paris-Saclay},
  addressline={ 4 Av. des Sciences, Gif-sur-Yvette},
  city={Paris},
  postcode={91190},
  country={France}
}

\affiliation[csrn]{
  organization={Center for Spintronics Research Network (CSRN), The University of Tokyo},
  addressline={ 7-3-1 Hongo, Bunkyo-ku},
  city={Tokyo},
  postcode={113-8656},
  country={Japan}
}

\affiliation[eeis]{
  organization={ Department of Electrical Engineering and Information Systems (EEIS), The University of Tokyo,},
  addressline={ 7-3-1 Hongo, Bunkyo-ku},
  city={Tokyo},
  postcode={113-8656},
  country={Japan}
}

\affiliation[aist]{
  organization={ Materials DX Research Center, National Institute of Advanced Industrial Science and Technology (AIST)},
  addressline={ 1-1-1 Umezono},
  city={Tsukuba},
  postcode={305-8568},
  country={Japan}
}

\affiliation[ias]{
  organization={ Institute for Advanced Study (IAS), Nagoya University},
  addressline={ Furo-cho, Chikusa-ku},
  city={Nagoya},
  postcode={464-8601},
  country={Japan}
}

\affiliation[imass]{
  organization={ Institute of Materials and Systems for Sustainability (IMASS), Nagoya University},
  addressline={ Furo-cho, Chikusa-ku},
  city={Nagoya},
  postcode={464-8601},
  country={Japan}
}

\affiliation[toyama]{
  organization={Electric and Electronic Engineering, Faculty of Engineering, University of Toyama},
  city={Toyama},
  postcode={930-8555},
  country={Japan}
}

\begin{abstract}
    We theoretically investigate the atomic-scale structure of the heterointerface formed between the (001) surface of the $L1_0$-ordered iron palladium (FePd) intermetallic alloy and graphene (Gr), namely, $L1_0$-FePd(001)/Gr, which serves as an essential component in spintronic devices.
    Using density functional theory (DFT) calculations, we demonstrate that the topmost surface layer consisting of Pd (Pd-terminated surface) is energetically more stable than that consisting of Fe in vacuum, and that Pd-terminated surfaces are unfavorable for graphene adsorption.
    In contrast, under an oxygen atmosphere, the strong Fe--O bonding stabilizes Fe-terminated surfaces.
    The predicted Fe--O bonds on the oxidized surface are consistent with our X-ray photoelectron spectroscopy (XPS) measurements.
    These results reproduce the mechanism responsible for the graphene coverage observed in recent experiments.
    Similar oxygen-induced Fe surface segregation has been studied in heterogeneous catalysis on FePt and FePd alloys.
    In this work, we exploit this mechanism as a termination-engineering strategy to fabricate high-quality 2D-material/alloy heterointerfaces for nanoscale device applications.
\end{abstract}

\begin{graphicalabstract}
\includegraphics[width=\columnwidth]{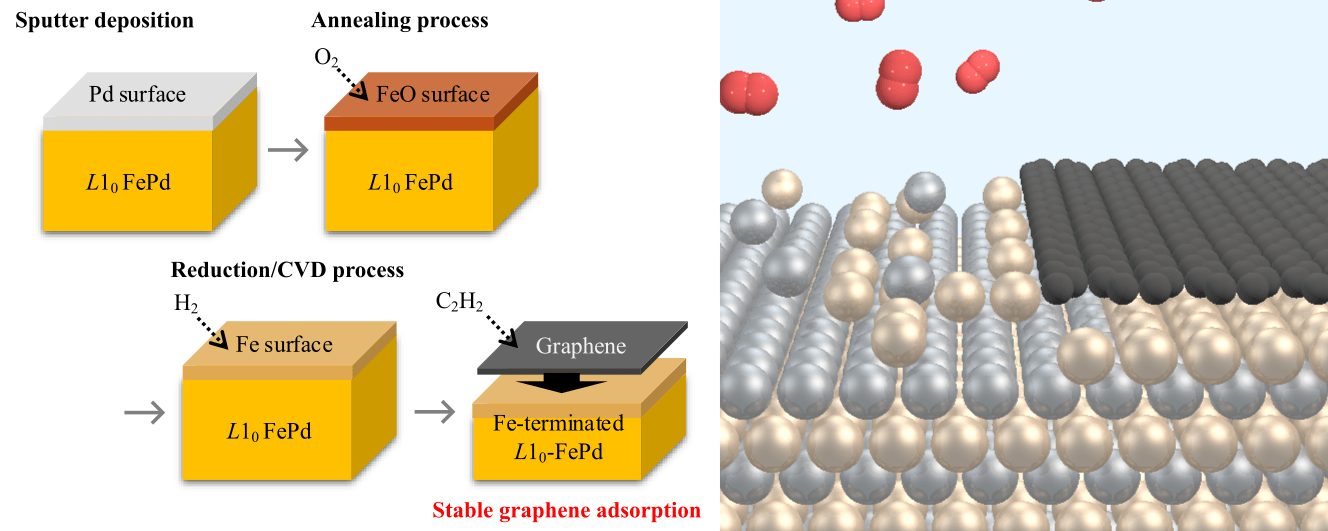}
\end{graphicalabstract}

\begin{highlights}
\item
DFT and XPS: Pd-terminated $L1_0$-FePd(001) in vacuum, Fe-terminated under oxygen.
\item
Oxidation--reduction during CVD yields graphene-covered Fe-terminated interfaces.
\item
Oxygen-induced segregation enables termination engineering of 2D/alloy interfaces.
\end{highlights}

\begin{keyword}
$L1_0$-ordered alloy \sep
Surface segregation \sep
Surface thermodynamics \sep
DFT

\end{keyword}

\end{frontmatter}

\section{\label{sec:intro} Introduction}

Iron palladium (FePd) \cite{naganuma2015electrical, naganuma2020perpendicular, naganuma2022unveiling, naganuma2023jpcc, uemoto2022density, adachi2024transport} has tetragonal $L1_0$-ordered crystal structures and exhibits anisotropic ferromagnetism.
FePd and similar binary alloys, such as FePt and FeIr, have large perpendicular magnetic anisotropy (PMA) \cite{itabashi2013preparation, shima2004lattice, klemmer1995magnetic, iihama2014low, kawai2014gilbert, zhang2018enhancement}, suggesting their potential use in spintronics applications such as magnetic storage devices including spin-transfer torque magnetoresistive random-access memory (STT-MRAM) \cite{naganuma2023spintronics, endoh2020recent, bhatti2017spintronics}.

The surface atomic composition of such Fe-based binary intermetallics has been investigated in the field of heterogeneous catalysis, where FePt- and FePd-based alloys and nanoparticles have attracted attention as catalysts for oxidation and gas-reforming-related reactions.
Experimental studies in this field have established that the terminating species of the alloy surface is highly sensitive to the surrounding chemical environment: under vacuum or reducing conditions, the noble-metal component (Pt or Pd) preferentially covers the surface, whereas exposure to even trace amounts of oxygen drives the segregation of Fe toward the surface accompanied by the formation of iron oxides \cite{han2009fe, prabhudev2015surface, li2020oxidation}.
For example, Han \textit{et al.} have shown by X-ray photoelectron spectroscopy that oxygen exposure of FePt nanoparticles and films selectively oxidizes Fe while segregated Pt layers protect the surface \cite{han2009fe}, and Prabhudev \textit{et al.} have observed that even trace oxygen at the ppm level induces Fe surface segregation in Pt--Fe nanoparticles \cite{prabhudev2015surface}.
More directly related to the present system, Li \textit{et al.} have reported that Fe atoms in a Pd--Fe alloy surface segregate to the topmost layer and form FeO structures under an O$_2$ pressure as low as $10^{-7}$~mbar \cite{li2020oxidation}.
Such environment-driven restructuring of the surface composition is now recognized as a general feature of bimetallic systems \cite{tao2008reaction}.
These experimental findings are consistent with theoretical calculations for $L1_0$-ordered alloys: Taniguchi \textit{et al.} have found that the Pt-terminated surface of $L1_0$-FePt is more stable than the Fe-terminated one in vacuum \cite{taniguchi2008theoretical}, and Dannenberg \textit{et al.} have reported similar stability of Pt-terminated surfaces in the $L1_0$ and $L1_1$ phases of CoPt and MnPt \cite{dannenberg2009surface}.
However, these insights into environment-dependent surface termination have been discussed almost exclusively from the viewpoint of catalytic activity.
Their implications for spintronics, in which the composition of the terminating layer governs the interfacial magnetic anisotropy \cite{hammar2022theoretical} and spin-dependent transport, remain largely unexplored.

For decades, heterointerfaces between such ferromagnetic alloys and two-dimensional (2D) materials, such as graphene (Gr), have been investigated as magnetic tunnel junctions (MTJs) \cite{hashmi2020graphene, robertson2023comparing}, which are essential components of STT-MRAM.
Previously, Naganuma~\textit{et al.} have reported the experimental synthesis of heterointerfaces between FePd and graphene (FePd/Gr) using chemical vapor deposition (CVD) techniques \cite{naganuma2020perpendicular}.
Scanning transmission electron microscopy (STEM) observations indicate the existence of atomically flat and uniform coverage \cite{naganuma2022unveiling}.
First-principles calculations have also been performed to study the atomic structure and the electronic and magnetic states \cite{uemoto2022density}, as well as spin transport characteristics that represent magnetic resistance (MR) performance \cite{adachi2024transport}.
The sandwich junction structure composed of FePd and multilayer graphene, referred to as "FePd/m-Gr/FePd," demonstrates a significant MR ratio, reaching the order of  $10^2~\%$.
However, the atomic composition of a bare or graphene-covered FePd surface remains inadequately understood.
Usually, an iron-based $L1_0$-ordered binary alloy consists of alternately stacked Fe and another metal along the $[001]$ axis; therefore, the surface termination problem discussed above directly applies to the FePd(001) surface, and the composition of the topmost layer is also essential in MRAM applications.
Our recent experimental and theoretical investigations indicate that the topmost layer of a graphene-covered FePd surface primarily consists of Fe atoms \cite{naganuma2020perpendicular, naganuma2022unveiling, naganuma2023jpcc, uemoto2022density, adachi2024transport}, which appears to contradict the noble-metal termination expected for bare $L1_0$-alloy surfaces in vacuum \cite{taniguchi2008theoretical, dannenberg2009surface}.
A related theoretical study has indicated that Fe termination becomes energetically favorable at the FePt/MgO heterointerface \cite{taniguchi2008theoretical}, suggesting that the stable termination of an $L1_0$ alloy can be reversed by the chemical environment of its surface.
We consider that the origin of these discrepancies lies in the chemical bonding energy between the topmost metallic atoms and the covering 2D material or oxygen in the atmosphere.

In this work, we carried out first-principles calculations to clarify the atomic-scale structures of the FePd/Gr heterointerface.
We prepared slab supercell models for both bare surfaces and heterointerfaces, and determined the formation energies of Fe- or Pd-terminated FePd surfaces and their interactions with graphene coverage.
Additionally, we analyzed the effects of oxidation on the formation energy of an Fe- or Pd-terminated surface in an oxygen-rich atmosphere.
The results indicate that the Pd-terminated bare surface is stable in vacuum, which is consistent with the behavior observed in other $L1_0$ alloys.
The analysis of surface region interactions suggests that the proximity between Fe and C contributes to energy stability.
Additionally, in an oxygen atmosphere, the strong chemical bonding between Fe and O stabilizes the Fe-terminated surface.
From these results, we hypothesize that the FePd/Gr surface observed in the experiment can be explained as follows: the Pd-rich surface of pure FePd is transformed into an Fe-terminated surface through exposure to atmospheric oxygen and thermal annealing during the formation of the graphene layer via CVD in a reducing gas atmosphere.
We believe that these findings will contribute to improving the fabrication quality of $L1_0$-alloy and 2D material heterointerfaces, paving the way toward advanced spintronics applications.

\section{Method}

\begin{figure}[t]
    \centering
    \includegraphics[width=0.45\textwidth]{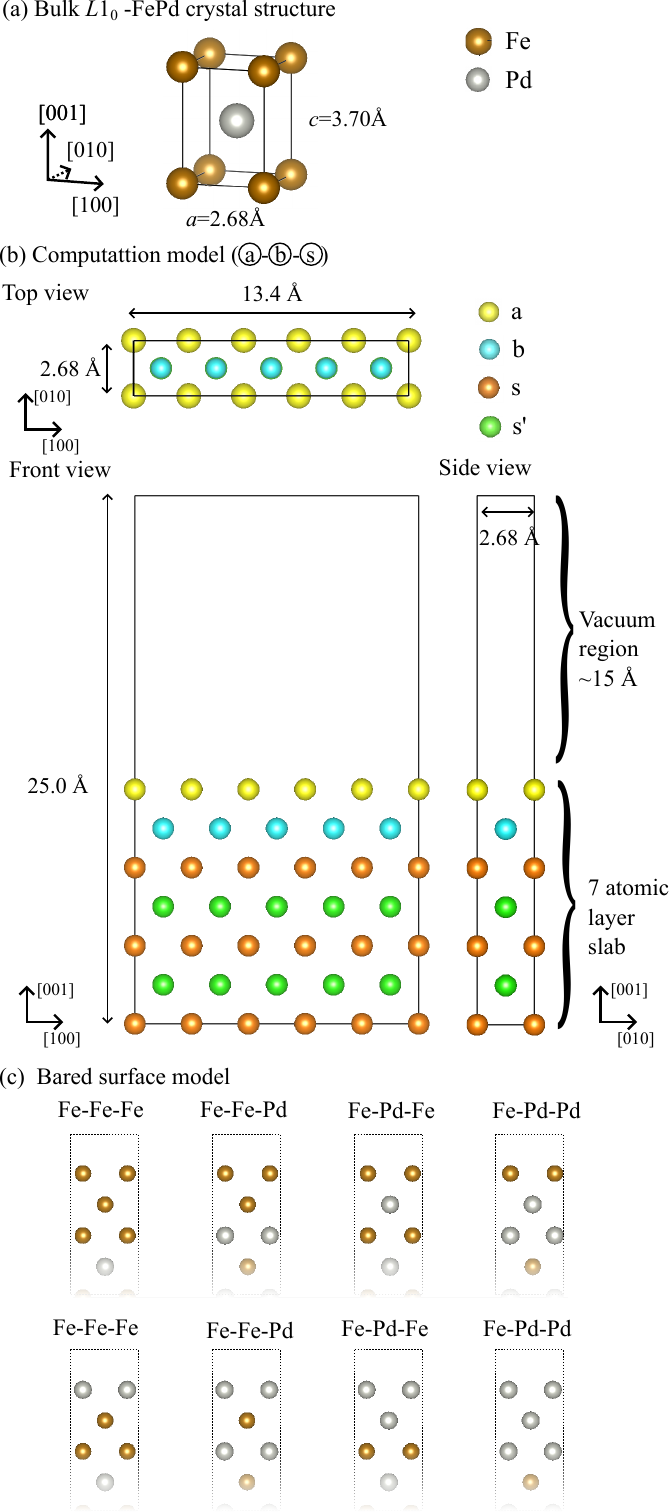}
    \caption{
        Schematic of the crystal structure and computational models.
        (a) Bulk $L1_0$-ordered FePd crystal cell under equilibrium conditions.
        (b) Slab supercell model of the pristine $L1_0$-FePd(001) surface labeled as $\textcircled{a}-\textcircled{b}-\textcircled{s}$, where \textcircled{a}, \textcircled{b}, \textcircled{s}, and \textcircled{s'} denote the Fe or Pd atoms in the topmost surface layer, the second layer, the odd-numbered substrate layers, and the even-numbered substrate layers, respectively.
        (c) Surface models with various atomic compositions.
    }
    \label{fig:model1}
\end{figure}

\begin{figure}[t]
    \centering
    \includegraphics[width=0.35\textwidth]{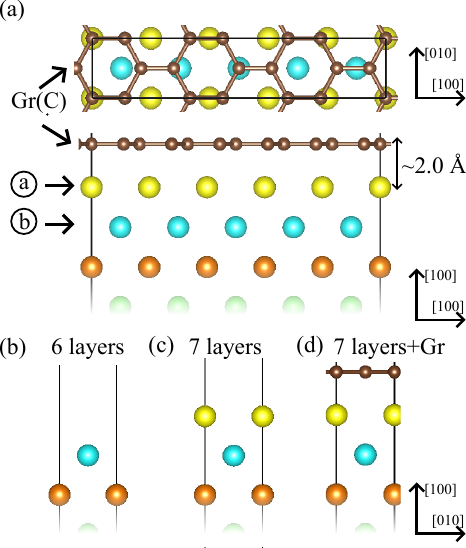}
    \caption{
        (a) Schematic of graphene-covered FePd surface.
        (b) Bare FePd surface model without the topmost atomic layer.
        (c) FePd surface model with the topmost atomic layer.
        (d) FePd surface model covered with the graphene layer.
    }
    \label{fig:model_stack}
\end{figure}

We performed first-principles calculations to investigate the stabilities of bare $L1_0$-FePd(001) surfaces and FePd/Gr heterointerfaces and analyzed the effect of the formation energy on structural configurations of surface metallic atoms.
Figure~\ref{fig:model1} illustrates our computational model.
The crystal structure of bulk $L1_0$-FePd is shown in Fig.~\ref{fig:model1}(a), and the equilibrium lattice constants ($a \approx 2.67~\text{\AA}$ and $c \approx 3.70~\text{\AA}$) were obtained from structural optimization \cite{uemoto2022density}.
In this work, we employed slab supercells that incorporate both vacuum and substrate regions.
Figure~\ref{fig:model1}(b) shows an overview of the slab supercell structure with dimensions of $13.35~\text{\AA} \times 2.67~\text{\AA} \times 26.28~\text{\AA}$.
The atomic structure consists of five periodic FePd units along the horizontal ($[100]$) direction.
In the vertical ($[001]$) direction, the slab consists of seven atomic layers with a total thickness of approximately $11.13~\text{\AA}$.
Each layer consists of either Fe or Pd atoms, denoted as \textcircled{a}, \textcircled{b}, \textcircled{s}, and \textcircled{s'}, representing atoms in the topmost surface layer, the second layer, the odd-numbered substrate layers, and the even-numbered substrate layers, respectively.
Notably, we assume that Fe and Pd layers are alternately stacked below the third layer, with \textcircled{s'} chosen to complement \textcircled{s}.
We considered several structural models with different surface metal compositions; for convenience, these models are labeled as ``\textcircled{a}--\textcircled{b}--\textcircled{s}'' (representing the elements in the first, second, and third layers, respectively).
Figure~\ref{fig:model1}(c) illustrates examples of several surface structures.
In addition, we also analyzed a heterointerface structure in which graphene is adsorbed onto the bare FePd(001) surface; the structure shown in Fig.~\ref{fig:model_stack}(a) has been proposed in our previous work, with further details provided in Ref.~\citenum{uemoto2022density}.

For computation, we used the Vienna \textit{ab initio} simulation package (VASP) code, which provides the first-principles electronic structure calculation based on the density functional theory (DFT) \cite{kresse1996efficiency}.
We used generalized gradient approximation (GGA) for exchange--correlation functionals \cite{perdew1996generalized} and the projector augmented wave (PAW) method for electron--ion interactions \cite{blochl1994projector}.
Only collinear spin polarization was considered in this calculation, and the effects of spin--orbit coupling were ignored.
Moreover, to account for van der Waals (vdW) interactions, we applied two widely used empirical functionals: Grimme's DFT-D2\cite{grimme2006semiempirical} and Klim\v{e}s' optB86b-vdW\cite{klimes2011}; the comparison of the results obtained with these functionals demonstrates the robustness of our calculations with respect to the choice of the vdW functional.
We used the $\Gamma$-centered Monkhorst--Pack $k$-point mesh with a $2 \times 10 \times 1$ grid or its equivalent, a plane wave cutoff energy of $500$~eV.
The convergence threshold for structural optimizations using the conjugate gradient method was set to be less than $10^{-3}$~eV, whereas the threshold for self-consistent field iterations for the electronic system was set to be less than $10^{-4}$~eV.

In addition, by modifying the slab supercell model described above, we also constructed graphene-covered and oxidized metal surface models; the details are provided in a later section.

\section{Results and discussion}

\subsection{Stability of alloy surface with graphene}

The formation energy to stack an additional single atomic layer on the surface is crucial for predicting stable structures.
Here, we calculated the formation energy for the topmost layer of FePd $E_\text{form}^\text{FePd}$ expressed as:
\begin{linenomath}
    \begin{align}
    E_\text{form}^\text{FePd}
    =&
    \left(
        E_\text{slab}^\text{w/\textcircled{a}}
        -
        E_\text{slab}^\text{w/o\textcircled{a}}
        -
        \mu_\text{Fe}
        N_\text{Fe}^\text{\textcircled{a}}
        -
        \mu_\text{Pd}
        N_\text{Pd}^\text{\textcircled{a}}
    \right)
    /
    N
    \;,
    \label{eq:e_form_fepd}
    \end{align}
\end{linenomath}
where $E_\text{slab}^\text{w/\textcircled{a}}$ and $E_\text{slab}^\text{w/o\textcircled{a}}$ represent the total energies of the bare FePd surface model with and without the topmost $\textcircled{a}$ layer, respectively [see Figs.~\ref{fig:model_stack}(b) and ~\ref{fig:model_stack}(c)].
$N_\text{Fe}^\text{\textcircled{a}}$ and $N_\text{Pd}^\text{\textcircled{a}}$ denote the numbers of Fe and Pd atoms in the topmost layer, respectively, and $N=N_\text{Fe}^\text{\textcircled{a}}+N_\text{Pd}^\text{\textcircled{a}}$.
$\mu_\text{Fe}$ and $\mu_\text{Pd}$ represent the chemical potentials of Fe and Pd, respectively.
Generally, chemical potentials depend on external conditions, such as material concentration.
Since the experimental growth conditions were reported to be Pd-rich \cite{naganuma2020perpendicular,naganuma2022unveiling}, according to a phase diagram of an Fe--Pd alloy \cite{burzo2010magnetic}, we assumed the coexistence of FePd and FePd$_3$ (or FePd and Pd). 
Within this environment, $\mu_\text{Fe}$ ranges from -8.63 to -8.75~eV, whereas $\mu_\text{Pd}$ ranges from -5.99 to -5.87~eV (DFT-D2); for further details, refer to Supplementary Material S1.

\begin{figure}[t]
    \centering
    \includegraphics[width=0.40\textwidth]{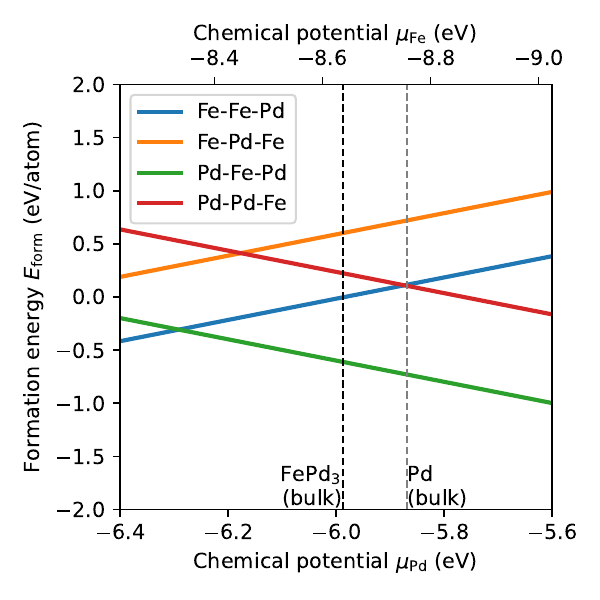}
    \caption{
        Formation energy $E_\text{form}^\text{FePd}$ from Eq.~(\ref{eq:e_form_fepd}) for various surface models:  Fe--Fe--Pd, Fe--Pd--Fe, Pd--Fe--Pd, and Pd--Pd--Fe.
        The black and gray dashed lines represent the chemical potentials of bulk FePd$_3$ and Pd, respectively.
        The calculations were performed using the DFT-D2 vdW functional.
    }
    \label{fig:surface_formation_energy}
\end{figure}

Figure~\ref{fig:surface_formation_energy} shows the formation energy $E_\text{form}^\text{FePd}$, calculated using Eq.~(\ref{eq:e_form_fepd}), for four different surface models: Fe--Fe--Pd, Fe--Pd--Fe, Pd--Fe--Pd, and Pd--Pd--Fe.
The black and gray dashed lines represent the chemical potentials of bulk FePd$_3$ and Pd, respectively.
Under these conditions, the results reveal the following order of stability: $\text{Pd--Fe--Pd} < \text{Fe--Fe--Pd} \sim \text{Pd--Pd--Fe} < \text{Fe--Pd--Fe}$.
The most stable structure has Pd as the topmost layer, which is consistent with theoretical studies on other Fe-based $L1_0$ binary alloys \cite{taniguchi2008theoretical,dannenberg2009surface}.

For simplicity, our analysis is limited to configurations where the atoms in the second and third layers are of different types. Additional configurations for the model shown in Fig.~\ref{fig:model1}(c) are summarized in Supplementary Material S2.

For the surface with the adsorbed graphene, we define the formation energy $E_\text{form}^\text{Gr}$ as:
\begin{linenomath}
\begin{align}
     E_\text{form}^\text{Gr} =&
     \left(
     E_\text{slab}^\text{w/~ Gr} -
     E_\text{slab}^\text{w/o Gr} -
     \mu_\text{C} N_\text{C}
     \right)
     /
     N
     \;,
\end{align}
\end{linenomath}
where $E_\text{slab}^\text{w/o Gr} $ and $ E_\text{slab}^\text{w/ Gr}$ represent the total energies of the slab surfaces without and with the graphene layer, respectively [see Fig.~\ref{fig:model_stack}(c) and \ref{fig:model_stack}(d)].
$\mu_\text{C}$ is the chemical potential of isolated pristine graphene, which is assumed to be $\mu_\text{C} \approx -9.28$~eV (DFT-D2).

In Table~\ref{tbl:e_form}, we summarize the calculated formation energies for four surface structures considered in Fig.~\ref{fig:surface_formation_energy}.
The values for $\mu_\text{Fe}$ and $\mu_\text{Pd}$ are taken from the black line in Fig.~\ref{fig:surface_formation_energy} (FePd$_3$).
For comparison, the results obtained using the DFT-D2 and optB86b functionals are provided (the values for optB86b are shown in parentheses).
From the $E^\text{FePd}_\text{form}$ values, in most cases, Pd-terminated surfaces are more stable than Fe-terminated surfaces.
For instance, stacking Fe layers on a Pd--Fe--Pd structure would result in an Fe--Pd--Fe surface. However, the formation energy indicates that the Pd--Pd--Fe surface is more stable.
Additionally, although Fe-terminated surfaces are energetically unstable (see $E_\text{form}^\text{FePd}$ for Fe--Pd--Fe), they can be stabilized by capping with a Pd layer (see $E_\text{form}^\text{FePd}$ for the Pd--Fe--Pd surface), resulting in the formation of the Pd--Fe--Pd structure.

\begin{table}[t]
    \centering
    \begin{tabular}{cccc}
        \hline
        &
       $E_\text{form}^\text{FePd}$ &
       $E_\text{form}^\text{Gr}$ &
       $E_\text{form}^\text{FePd}+E_\text{form}^\text{Gr}$ \\
        &
        [eV/atom] &
        [eV/atom] &
        [eV/atom] \\
        \hline
        Fe-Fe-Pd & +0.00 & -0.04 & -0.05  \\
            & (+0.03) & (-0.02) & (+0.01) \\
        \hline
        Fe-Pd-Fe & +0.60 & -0.21 & +0.39  \\
            & (+0.59) & (-0.15) & (+0.44) \\
        \hline
        Pd-Fe-Pd & -0.61 & +0.24 & -0.37  \\
            & (-0.59) & (+0.32) & (-0.28) \\
        \hline
        Pd-Pd-Fe & +0.22 & +0.15 &  +0.37 \\
            & (+0.21) & (+0.24) & (+0.45) \\
        \hline
    \end{tabular}
    \caption{
        \label{tbl:e_form}
        Calculated formation energies of bare FePd surfaces $E_\text{form}^\text{FePd}$ and $E_\text{form}^\text{Gr}$ for various different surface structure models in Fig.~\ref{fig:model1}.
        For comparison, the results using the DFT-D2 and optB86b functionals are provided (the values for optB86b are shown in parentheses).
    }
\end{table}

As shown in Table~\ref{tbl:e_form}, the formation energies ($E_\text{form}^\text{Gr}$) of the structures follow the order $\text{Fe--Pd--Fe} < \text{Fe--Fe--Pd} < \text{Pd--Fe--Pd} < \text{Pd--Pd--Fe}$, indicating that the bonding between Fe and C is significantly stronger than that between Pd and C.
A similar trend has been reported in theoretical calculations for interfaces between other iron-based alloys and graphene, where the stabilization is originating to the hybridization of the unoccupied Fe minority-spin $d$ orbitals with the C $p_z$ orbitals with charge transfer \cite{matsumoto2025theoretical}.
Therefore, the bare Fe-terminated surface, which is initially unstable, can be stabilized by capping with a graphene layer (or other metallic layers).
However, the gain in $E_\text{form}^\text{Gr}$ is smaller in magnitude than that in $E_\text{form}^\text{FePd}$.
For the total formation energy, $E_\text{form}^\text{FePd} + E_\text{form}^\text{Gr}$ follows the order $\text{Pd--Fe--Pd} < \text{Fe--Fe--Pd} < \text{Fe--Pd--Fe} \approx \text{Pd--Pd--Fe}$; the Pd-terminated surface with adsorbed Gr is still the most stable.
This result contrasts with those in Ref.~\citenum{naganuma2022unveiling}, which indicates the existence of the Fe-terminated surface with adsorbed Gr; the attractive chemical interaction between Fe and C is insufficient to explain this discrepancy.

\subsection{Effect of surface oxidation}
In the previous section, we discussed the stability of FePd/Gr in a vacuum and suggested that the formation of Fe-terminated surfaces is unlikely under such conditions.
Next, we additionally explore the effect of surface oxidation as a potential mechanism for forming Fe-terminated surfaces.
Under the experimental conditions, FePd samples were fabricated using the sputtering method \cite{naganuma2020perpendicular,naganuma2022unveiling}, and the surface was exposed to an oxygen atmosphere before the deposition of graphene by CVD.
To investigate the effect of oxygen, we performed X-ray photoelectron spectroscopy (XPS) measurements on FePd samples.
XPS spectra were obtained using a Mg-$K\alpha$ X-ray source ($h\nu = 1253.6$~eV); photoelectrons were collected using a Gammadata Scienta SES-100 hemispherical analyzer in transmission mode.
The measurements were conducted at room temperature at a base pressure of approximately $1.0 \times 10^{-7}$~Pa.

\begin{figure*}[htbp]
    \centering
    \includegraphics[width=0.80\textwidth]{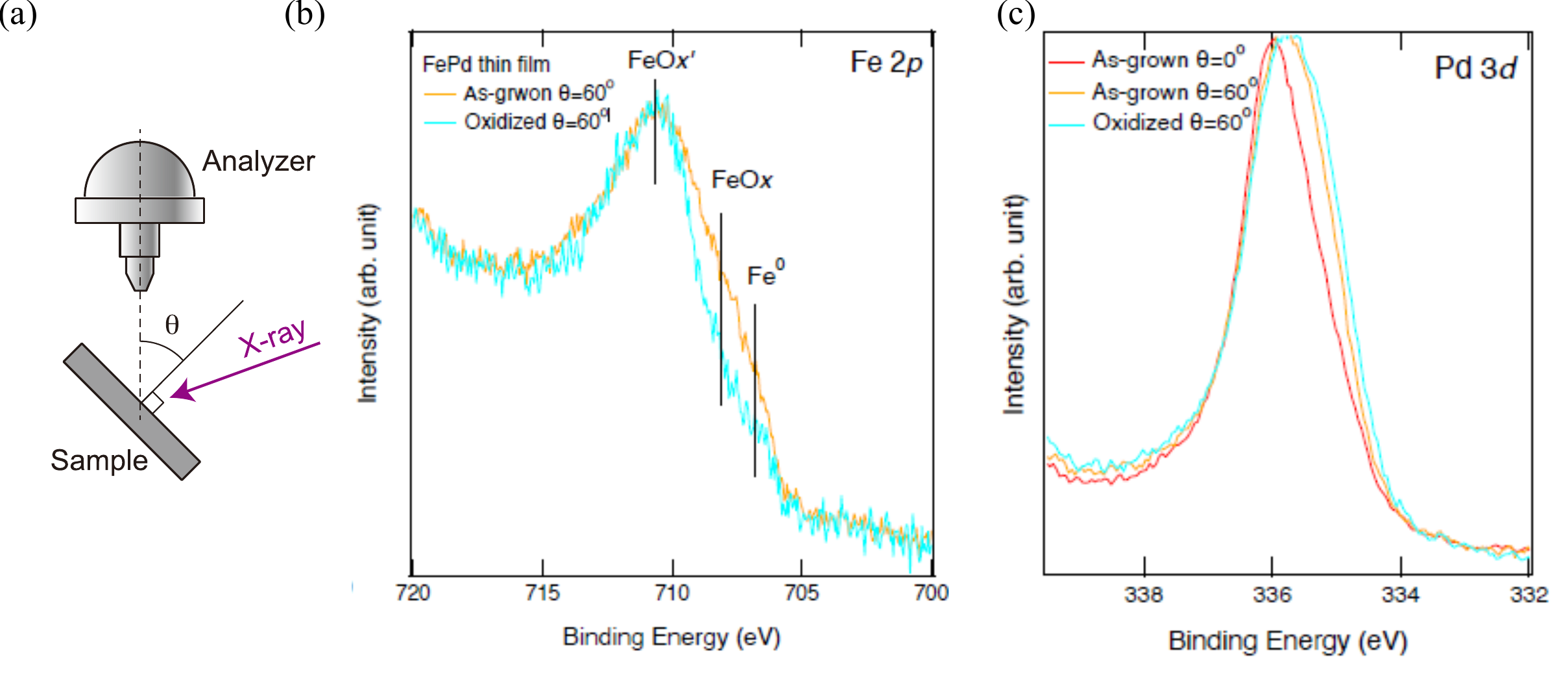}
    \caption{
        Schematic of the experimental conditions (a) and spectra at the Fe~$2p$ (b) and Pd~$3d$ (c) core levels.
        The spectra for samples immediately after fabrication are shown in yellow, and those after one week of oxidation in the air are shown in light blue.
    }
    \label{fig:xps_spectra}
\end{figure*}

Figure~\ref{fig:xps_spectra} presents the schematic of the experimental setup and results of XPS measurement at Fe~$2p$ and Pd~$3d$  peaks.
We compared the spectra of samples immediately after fabrication (labeled ``As-grown'') and after one week of oxidation in the air (labeled "Oxidized").
The samples under the latter condition closely resemble the samples used in prior FePd/Gr fabrication\cite{naganuma2020perpendicular,naganuma2022unveiling}.
In Fig.~\ref{fig:xps_spectra}(b), the Fe~$2p$ core-level spectrum indicates the increase in the intensity of the FeO$_x$ peak in the oxidized sample; it indicates the existence of bonding between Fe and O.
In Fig.~\ref{fig:xps_spectra}(c), it can be observed that Pd atoms are less susceptible to oxidation under the same conditions compared with Fe atoms.
This selective oxidation of Fe over the noble-metal component agrees with previous XPS studies of FePt nanoparticles and thin films \cite{han2009fe} as well as with the oxidation-induced segregation of FeO observed on Pd--Fe alloy surfaces \cite{li2020oxidation}.

\begin{figure*}[htbp]
    \centering
    \includegraphics[width=0.80\textwidth]{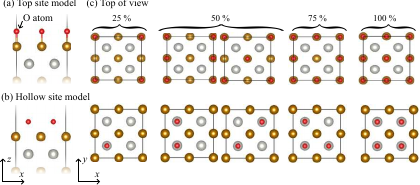}
    \caption{
        \label{fig:oxidation_model}
        Computational models of oxidized FePd surfaces (Fe-terminated case). Two types of adsorption sites are considered: top (a) and hollow (b).
        Various oxygen coverage ratios and configurations are illustrated in (c).
    }
\end{figure*}

Next, we performed first-principles calculation to evaluate the effect of the oxidization of surface metal atoms on the formation energy; we constructed a model that includes O atoms, as illustrated in Fig.~\ref{fig:oxidation_model}.
The atomic scale configuration of O atoms on the FePd(001) surface is as yet not well understood. Therefore, we propose several initial structural models based on the well-known crystals of FeO and PdO (with a rocksalt structure).
In this study, we consider two adsorption sites (referred to as ``top'' and ``hollow'') and four different coverage rates: 25, 50, 75, and 100~\%, where 100~\% indicates a completely oxidized surface (the ratio of topmost metal to oxygen is 1:1).
Examples of Fe-terminated surface structures are shown in Fig.~\ref{fig:oxidation_model}(c), and the corresponding Pd-terminated surface structures can be obtained.
To obtain the most realistic formation energies of the oxidized surfaces, structural optimization was performed for each proposed model.
The formation energy ($E^\text{O}_\text{form}$) is expressed as:
\begin{linenomath}
\begin{align}
E^\text{FePd/O}_\text{form}
=& E^\text{FePd}_\text{form} + E^\text{O}_\text{form}
\label{eq:e_form_o}
\end{align}
\end{linenomath}
with
\begin{linenomath}
\begin{align}
    E^\text{O}_\text{form}
    =&
    (E^\text{w/ O}_\text{surf} - E^\text{w/o O}_\text{surf} - N_\text{O} \mu_\text{O}  )
    /N
    \label{eq:e_form_o_oxygen}
    \;,
\end{align}
\end{linenomath}
where $E^\text{FePd}_\text{form}$ is the formation energy of the topmost metallic layer as defined in Eq.~(\ref{eq:e_form_fepd}).
$E^\text{w/ O}_\text{surf}$ represents the energy of the oxidized FePd slab and $E^\text{w/o O}_\text{surf}$ corresponds to the energy of the bare surface. $N$ is the number of atoms in the topmost layer, $N_\text{O}$ is the number of O atom, and $\mu_\text{O}$ is the chemical potential of oxygen from diatomic molecule in atmosphere, which is given as below:
\begin{linenomath}
\begin{align}
 \mu_\text{O}(p_{\text{O}_2}, T)
 =&
 \frac{1}{2} \Big[
 E_{\text{O}_2}
 +
 k_\text{B} T
 \log \left( \frac{p_{\text{O}_2}}{p^\circ} \right)
 \notag \\ &
 +
 \Delta_\text{f} G_{\text{O}_2}(p^\circ, T)
 -
 \Delta_\text{f} G_{\text{O}_2}(p^\circ, 0)
 \Big]
 \;,
\end{align}
\end{linenomath}
where $E_{\text{O}_2}$ is the energy of an isolated $\text{O}_2$ molecule.
Here, $p_{\text{O}_2} / p^\circ$ represents the relative partial pressure of $\text{O}_2$ gas and $\Delta_\text{f} G_{\text{O}_2}$ is the Gibbs free energy for a single molecule \cite{chase1998}.

As a result, in the case of Fe-terminated surfaces, the formation energy of oxidation ($E_\text{form}^\text{O}$) is minimized at a hollow site with a coverage rate of 100~\%, which reaches $E_\text{form}^\text{O} \approx -2.85$~eV/atom.
This energy gain is significantly larger than the formation energy of the bare surface ($E_\text{form}^\text{FePd}$); then, the total formation energy $E_\text{form}^\text{FePd/O} \approx -2.25$~eV/atom. 

\begin{figure}[t]
    \centering
    \includegraphics[width=0.40\textwidth]{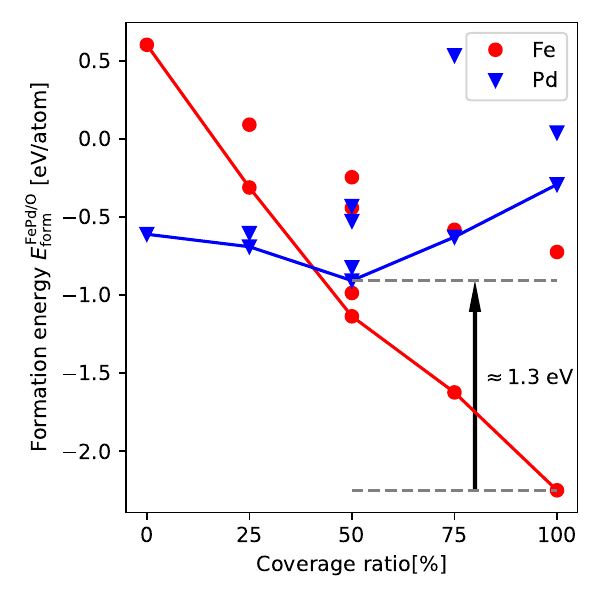}
    \caption{
        \label{fig:oxidation_energy}
        Calculated formation energy $E_\text{form}^\text{FePd/O}$ as defined in Eq.~(\ref{eq:e_form_o}) plotted as a function of the oxygen coverage ratio for Fe-terminated (red) and Pd-terminated surface models.
        The solid line indicates the minimum $E_\text{form}^\text{FePd/O}$ (convex hull) at each ratio.
    }
\end{figure}

We performed similar calculations for Pd-terminated surface models.
Figure~\ref{fig:oxidation_energy} illustrates the calculated formation energy $E_\text{form}^\text{FePd/O}$ as a function of the oxygen coverage ratio.
The minimum $E_\text{form}^\text{FePd/O}$ of the Fe-terminated surface models is approximately 1.3~eV lower than that of the Pd-terminated surface models.
In contrast to the analysis results in above section,
these results suggest that the stability of Fe surfaces are improved in samples exposed to an oxygen atmosphere.
This finding is consistent with the results obtained from XPS measurements in Fig.~\ref{fig:xps_spectra} and reproduces the prior characterization of the FePd/Gr interface by STEM \cite{naganuma2022unveiling}.
It also parallels the oxygen-induced Fe surface segregation established experimentally for FePt- and Pd--Fe-based alloys in the catalysis literature \cite{han2009fe, prabhudev2015surface, li2020oxidation}.

\subsection{Formation mechanism}
\begin{figure*}[htbp]
    \centering
    \includegraphics[width=16cm]{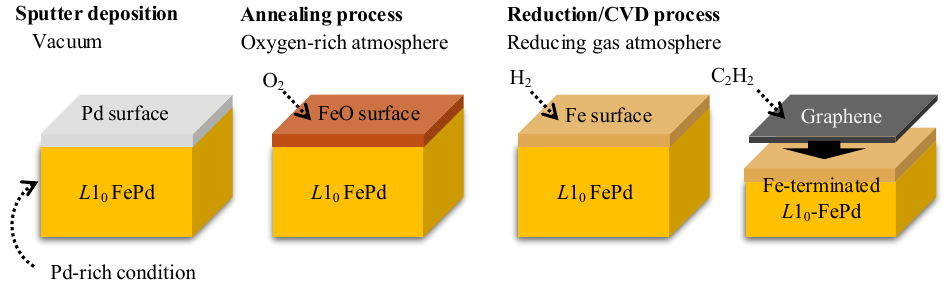}
    \caption{\label{fig:formation_mechanism}
        Proposed formation mechanism of FePd/Gr
    }
\end{figure*}

In the analysis presented 
above,  
we investigated the stability of the FePd surface in vacuum.
We determined that Pd-terminated surfaces would be energetically more favorable than Fe.
Moreover, the adsorbed graphene is unstable in the case of the Pd-terminated surface.
Then, the most energetically favorable configuration becomes the Pd surface without graphene.
In the analysis in the previous section, 
we established that the formation mechanism of the Fe-terminated surface, promoted by the presence of oxidized iron on the FePd surface under an oxygen atmosphere, enhances the thermodynamic stability.
This finding is in agreement with the results obtained from XPS measurements.
From the findings above, we proposed a formation mechanism for the experimentally observed FePd/Gr heterointerface, as shown in Fig.~\ref{fig:formation_mechanism}.

In the experimental fabrication process, $L1_0$-FePd is grown by radio frequency magnetron sputtering using a FePd target with a 46:54 atomic ratio, resulting in a FePd film with a 50:50 atomic ratio.
The FePd films are annealed at high temperatures and temporarily exposed to the atmosphere before being transferred to the CVD chamber.
During CVD, the sample is annealed under a reducing gas atmosphere (H$_2$) to remove the oxidized surface layer, and graphene is deposited on the surface using C$_2$H$_2$ gas (the details are given in Ref.~\citenum{naganuma2022unveiling}).

As illustrated in Fig.~\ref{fig:formation_mechanism}, we hypothesize that the initially Pd-rich surface of the sample becomes covered with FeO owing to oxidation.
In the subsequent stage, the hydrogen-reduced Fe surface provides favorable sites for carbon adsorption.
This process likely enables the formation of a macroscale graphene-covered Fe-terminated surface.
We note that such a compositional reconstruction driven by sequential oxidizing and reducing atmospheres is analogous to the reaction-driven restructuring of bimetallic catalysts \cite{tao2008reaction, li2020oxidation}; the proposed mechanism thus exploits, at a spintronics interface, the same segregation chemistry that has been established in heterogeneous catalysis.
Additionally, we consider that the above does not preclude the formation of the energetically stable bare (uncovered) Pd-terminated surface;
under specific growth conditions, samples with a higher proportion of Pd-terminated surfaces could also be obtained.

\subsection{Interfacial electronic and magnetic structure}
\begin{figure*}[htbp]
    \centering
    \includegraphics[width=0.7\textwidth]{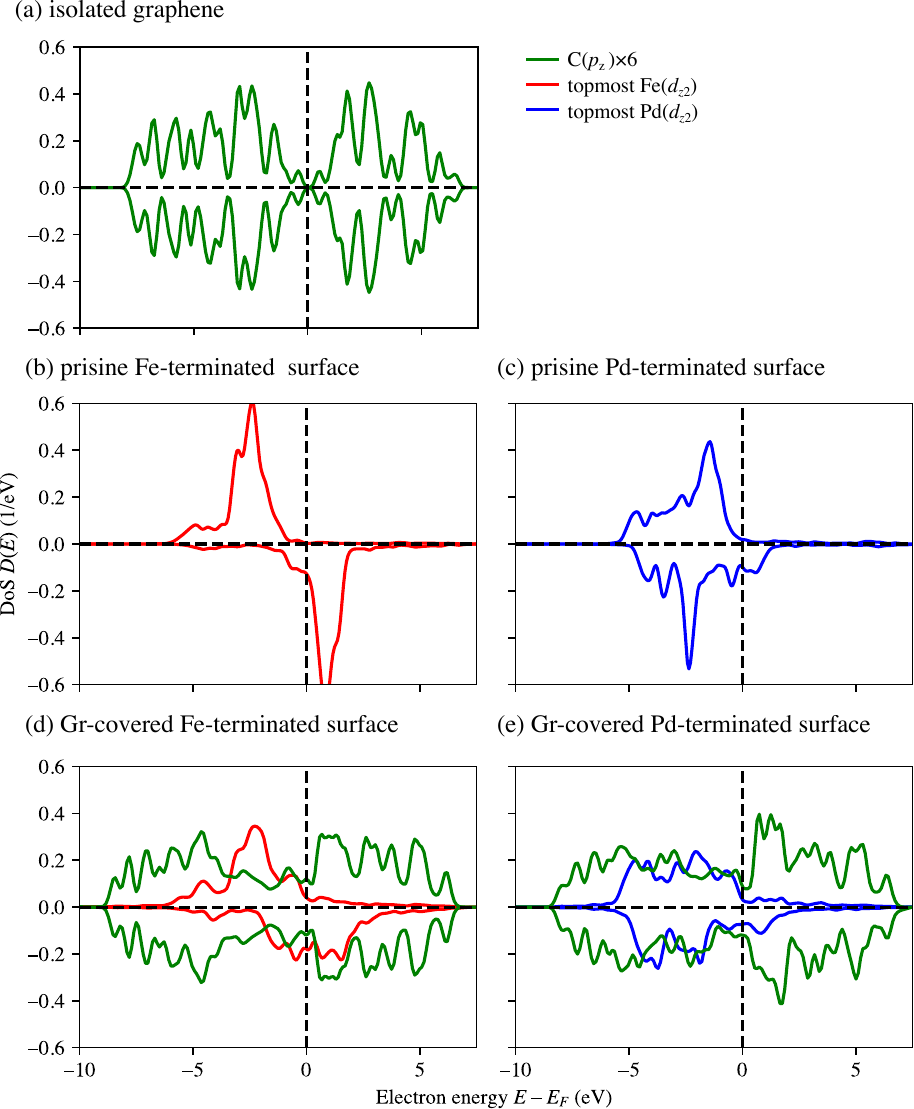}
    \caption{
        \label{fig:dos}
        Electronic density of states: $p_z$ orbitals of isolated graphene (a); $d_{z^2}$ orbitals of the topmost metallic layer on the pristine Fe-terminated surface (b) and on the pristine Pd-terminated surface (c); and on the graphene-covered Fe-terminated surface (d) and Pd-terminated surface (e).
}
\end{figure*}
Finally, we examined the electronic states at the interface. Figure~\ref{fig:dos} shows the partial density of states (pDoS). For isolated graphene, the $p_z$ orbital contributes states near the Fermi level [Fig.~\ref{fig:dos}(a)].

For the topmost metallic layer of the pristine surfaces, the $d_{z^2}$ orbitals of the Fe- and Pd-terminated surfaces are shown in Figs.~\ref{fig:dos}(b) and (c), respectively. For Fe, the exchange splitting is complete, and the minority-spin $d_{z^2}$ orbital is almost unoccupied, whereas for Pd the splitting of the $d_{z^2}$ states is minimal. Accordingly, the magnetic moment of the surface Fe atoms reaches $3.0~\mu_{\mathrm{B}}$, while that of the Pd atoms is only $0.3~\mu_{\mathrm{B}}$.

When the surface is covered by graphene, bonding states emerge between the $p_z$ orbitals of graphene and the metal $d_{z^2}$ orbitals \cite{uemoto2022density}.
For the Fe-terminated surface [Fig.~\ref{fig:dos}(d)], the minority-spin bands hybridize with these bonding states and become partially occupied, reflecting an electron transfer from the graphene $p_z$ states into the Fe minority-spin $d_{z^2}$ channel.
Similar behavior has been reported at other iron-based alloy--graphene heterointerfaces \cite{matsumoto2025theoretical}.
As a result, the average magnetic moment of Fe decreases to $2.6~\mu_{\mathrm{B}}$, and the surface C atoms acquire a spin polarization with a magnetic moment of approximately $-0.02~\mu_{\mathrm{B}}$. 
By contrast, for the Pd-terminated surface [Fig.~\ref{fig:dos}(e)], the change of magnetic moment are relatively small: magnetic moment of the Pd layer still remains $\sim 0.3~\mu_{\mathrm{B}}$, and the induced magnetic moment of C is nearly negligible, an order of magnitude smaller than in the Fe case.

Our results indicate that the interfacial electronic states and magnetism are highly sensitive to the topmost metal element, which in turn governs the spin-dependent tunneling transport essential to spintronic device operation \cite{adachi2024transport}; precise control of the terminating element during device fabrication is therefore indispensable.

\section{Conclusions}

\label{sec:summary}
In this study, we present a first-principles analysis of the van der Waals heterointerface between 2D materials and ferromagnetic alloys, namely FePd/Gr.
Our calculations show that, in vacuum, the Pd-terminated surface of bare FePd is the most stable, consistent with other $L1_0$ intermetallics.
Although graphene coverage promotes Fe termination through the attractive Fe--C bonding, this energy gain alone is insufficient to overcome the stability of the Pd-terminated surface.
Under an oxygen atmosphere, the formation of Fe--O bonds shifts the stability toward Fe termination; this selective oxidation of Fe is also observed in our XPS measurements.
On this basis, we propose a formation mechanism in which the initially Pd-rich surface is oxidized to FeO and subsequently reduced during chemical vapor deposition (CVD), yielding the graphene-covered Fe-terminated interface observed in previous STEM experiments \cite{naganuma2022unveiling}.

In addition, atmosphere-dependent surface segregation has been widely reported in heterogeneous catalysis \cite{han2009fe, prabhudev2015surface, li2020oxidation, tao2008reaction}.
We suggest that it can be utilized as a termination-engineering strategy applicable to Fe-based $L1_0$ intermetallic compounds, such as FePt and FeNi, to enable the selective realization of Fe- or noble-metal-terminated interfaces.
Because the terminating element governs the interfacial electronic structure and magnetism, and thereby the spin-dependent tunneling transport \cite{adachi2024transport}, such termination engineering provides a direct means of tailoring the performance of spintronic devices.
Our theoretical and experimental findings advance fabrication strategies for $L1_0$-alloy and 2D-material heterointerfaces and provide valuable insights for the development of future spintronic technologies.

\section*{Acknowledgement}
This work was partially financially supported by MEXT as part of the Japan Society for the Promotion of Science (JSPS) KAKENHI (24H01196, 24K01346) and Core-to-Core Program (No. JPJSCCA20230005) by Cooperative Research Project from CSRN, and by the cross-appointment project (H.N, P.S., and J.R.) and QST-Tohoku University matching foundation.
In addition, this work was also partially financially supported by MEXT as part of the ``Program for Promoting Researches on the Supercomputer Fugaku'' (Quantum-Theory-Based Multiscale Simulations toward the Development of Next-Generation Energy-Saving Semiconductor Devices, JPMXP1020200205), JST CREST(JPMJCR22B4), Kurata Grants, the Iwatani Naoji Foundation and the Spintronics Research Network of Japan (Spin-RNJ).
H.N. was partly supported by the Nagoya University Program for Research Enhancement.
The numerical calculations were carried out using the computer facilities of the Institute for
Solid State Physics at The University of Tokyo, the Center for Computational Sciences at the University of Tsukuba (Multidisciplinary Cooperative Research Program), and the supercomputer Fugaku provided by the RIKEN Center for Computational Science (Project ID: hp240178, hp250193, hp260170).
This research also used computational resources of (Wisteria/BDEC-01 and Miyabi) provided by Multidisciplinary Cooperative Research Program in Center for Computational Sciences, University of Tsukuba.
The visualization is performed by VESTA code~\cite{momma2008vesta}.
The authors gratefully acknowledge the assistance of ChatGPT (OpenAI) in checking the English language and grammar of the manuscript.

\section*{Data availability}
Data will be made available on request.

\bibliographystyle{elsarticle-num} 
\bibliography{refs.bib}

\end{document}


\maketitle
\footnotetext[1]{
    Department of Electrical and Electronic Engineering, 
    Graduate School of Engineering, 
    Kobe University, 
    1-1 Rokkodai-cho, Nada-ku, Kobe 651-8501, Japan
}
\footnotetext[2]{
    ENS Paris-Saclay, 
    4 Av. des Sciences, Gif-sur-Yvette, 91190, France
}
\footnotetext[3]{
    Center for Spintronics Research Network (CSRN),
    The University of Tokyo,
    7-3-1 Hongo, Bunkyo-ku, Tokyo 113-8656, Japan
}
\footnotetext[4]{
    Department of Electrical Engineering and Information Systems (EEIS),
    The University of Tokyo,
    7-3-1 Hongo, Bunkyo-ku, Tokyo 113-8656, Japan
}
\footnotetext[5]{
    Materials DX Research Center,
    National Institute of Advanced Industrial Science and Technology (AIST),
    1-1-1 Umezono, Tsukuba, Ibaraki 305-8568, Japan
}
\footnotetext[6]{
    Institute for Advanced Study (IAS), 
    Nagoya University, 
    Furo-cho, Chikusa-ku, Nagoya, 464-8601, Japan
}
\footnotetext[7]{
    Institute of Materials and Systems for Sustainability (IMaSS), 
    Nagoya University, 
    Furo-cho, Chikusa-ku, Nagoya, 464-8601, Japan
}
\footnotetext[8]{
    Electric and Electronic Engineering, Faculty of Engineering, University of Toyama,
    Toyama, 930-8555, Japan
}

\renewcommand{\thesection}{S.\arabic{section}}
\renewcommand{\thefigure}{S.\arabic{figure}}
\renewcommand{\theequation}{S.\arabic{equation}}
\renewcommand{\thetable}{S.\arabic{table}}

\clearpage
\section{Chemical potentials of Fe and Pd atoms}

The chemical potential of iron and paradium $\mu_\text{Fe}$ and $\mu_\text{Pd}$ is indeed depends on environmental properties such as the compotion of the alloy.
During the experimental synthesis of FePd in Pd-rich growth condition, based on the phase diagram of Fe--Pd system, the coexistence of multiple phases, such as  FePd and FePd$_3$ and Pd, can be achieved.
When two alloys in equilibrium (Alloy~A and Alloy~B), the chemical potential of a metallic components must be equal. Then we have:
\begin{align}
    \mu^\text{A}_\text{Fe} =& \mu^\text{B}_\text{Fe} \label{eq:mu1_fe}\\
    \mu^\text{A}_\text{Pd} =& \mu^\text{B}_\text{Pd} \label{eq:mu1_pd}
    \;.
\end{align}
The density functional theory (DFT) calculation provides the total energy $E$;  which is related to chemical potential by the following equation:
\begin{align}
    E(\text{A}) =& N^\text{A}_\text{Fe} \mu^\text{A}_\text{Fe} + N^\text{A}_\text{Pd} \mu^\text{A}_\text{Pd}
    \;,
    \label{eq:mu2_a}
\end{align}
where $N^\text{A}_\text{Fe}$ and $N^\text{A}_\text{Pd}$ are the numbers of Fe and Pd atoms in the unit cell of alloy A, respectively.
Besides, a similar relationship can be hold for the case of alloy B:
\begin{align}
    E(\text{B}) =& N^\text{B}_\text{Fe} \mu^\text{B}_\text{Fe} + N^\text{B}_\text{Pd} \mu^\text{B}_\text{Pd}
    \;.
    \label{eq:mu2_b}
\end{align}
Then, Eqs.~(\ref{eq:mu1_fe})-(\ref{eq:mu2_b}) lead to system of equations which provides the chemical potentials.

\begin{figure}[htbp]
    \centering
    \includegraphics[width=0.4\textwidth]{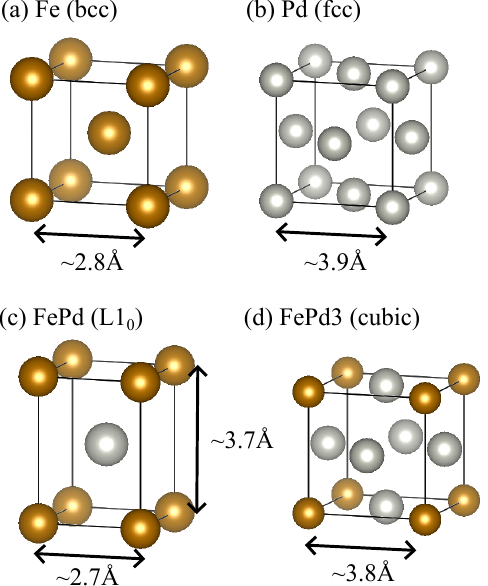}
    \caption{
        \label{fig:crystal_model}
        Crystal structures of (a) Fe, (b) Pd, (c) FePd, (d) FePd$_3$.
    }
\end{figure}

Here, we consider the a few reference materials: FePd, FePd$_3$, Pd (and Fe, for comparison), whose crystal structure are illustrated in Fig.~\ref{fig:crystal_model}.
The number of atoms in unit cell and corresponding total energies are listed in Table.~\ref{tbl:total_energy}.
The calculated $\mu_\text{Fe}$ and $\mu_\text{Pd}$ for each of the two alloys systems are also provides in Table.~\ref{tbl:chemical_potential}.
$\mu_\text{Fe}$ and $\mu_\text{Pd}$ are distributing in a narrow range of 0.3 eV regardress of the material system .

Additionally, for comparison, we employ two types of empirical van der Waals (vdW) functionals: DFT-D2 and optB86b-vdW.

\begin{table*}[h]
    \centering
    \caption{\label{tbl:total_energy}
        Crystal structures,number of atoms per unit cell, and calculated total energies for the considered reference materials obtained by two vdW functionals: DFT-D2 and optB86b-vdW.
    }
    \begin{tabular}{llcccc}
    \toprule
    \multicolumn{2}{l}{Material} & \multicolumn{2}{l}{Number of atoms [1/cell]} & \multicolumn{2}{l}{Total energy [eV/cell]} \\ 
    \cmidrule(lr){1-2} \cmidrule(lr){3-4} \cmidrule(lr){5-6}
    Name   & Structure           & $N_\text{Fe}$     & $N_\text{Pd}$     & DFT-D2               & optB86b-vdW               \\ \midrule
    Fe     & bcc                 & 2                 & 0                 & -17.35               & -11.91                    \\
    Pd     & fcc                 & 0                 & 4                 & -23.47               & -9.84                     \\
    FePd   & tetragonal($L1_0$)  & 1                 & 1                 & -14.63               & -8.46                     \\
    FePd3  & cubic               & 1                 & 3                 & -26.60               & -13.61                    \\ \bottomrule
    \end{tabular}
    \label{tab:total_energy}
\end{table*}

\begin{table*}[h]
    \centering
    \caption{\label{tbl:chemical_potential}
        Calculated chemical potential $\mu_\text{Fe}$ and $\mu_{Pd}$ from various materials.
    }
    \begin{tabular}{llcccc}
        \toprule
                     &                 & \multicolumn{4}{c}{Chemical potentials [eV]} \\
        \cmidrule{3-6}
        \multicolumn{2}{c}{Material}   & \multicolumn{2}{c}{DFT-D2 results} & \multicolumn{2}{c}{optB86b-vdw results} \\
        \cmidrule(r{4pt}){1-2} \cmidrule{3-4} \cmidrule{5-6}
        A           & B              & $\mu_\text{Fe}$  & $\mu_\text{Pd}$ & $\mu_\text{Fe}$     & $\mu_\text{Pd}$     \\
        \midrule
        Fe          & Pd             & -8.67            & -5.87           & -5.96               & -2.46               \\
        Fe          & FePd           & -8.67            & -5.95           & -5.96               & -2.51               \\
        Fe          & FePd$_3$       & -8.67            & -5.98           & -5.96               & -2.55               \\
        Pd          & FePd           & -8.76            & -5.87           & -6.00               & -2.46               \\
        Pd          & FePd$_3$       & -8.99            & -5.87           & -6.23               & -2.46               \\
        FePd        & FePd$_3$       & -8.64            & -5.99           & -5.89               & -2.57               \\
        \bottomrule  
    \end{tabular}
    \label{tab:chemical_potential}
\end{table*}

\clearpage
\section{Formation energies of bare and Gr-covered surfaces}

\begin{table}[htbp]
    \centering
    \caption{\label{tbl:chemical_potential}
        Formation energies of bare FePd surface $E_\text{form}^\text{FePd}$ and the adsorption of Gr layer $E_\text{form}^\text{Gr}$, which are defined in Sec.~2 of the manuscript.
    }
    \begin{tabular}{lcccccc}
    \toprule
    & \multicolumn{6}{c}{Formation energy [eV]} \\
    \cmidrule{2-7}
    & \multicolumn{3}{c}{DFT-D2 results} & \multicolumn{3}{c}{OptB86b-vdw   results} \\ 
    \cmidrule(lr){2-4} \cmidrule(lr){5-7}
    Model 
    & $E_\text{form}^\text{FePd}$ & $E_\text{form}^\text{Gr}$ & $E_\text{form}^\text{FePd}+E_\text{form}^\text{Gr}$ 
    & $E_\text{form}^\text{FePd}$ & $E_\text{form}^\text{Gr}$ & $E_\text{form}^\text{FePd}+E_\text{form}^\text{Gr}$ \\
    \midrule   
    Fe-Fe-Fe &	-0.03 &	-0.03 &	-0.06 &	-0.03 &	-0.01 &	-0.04 \\
    Fe-Fe-Pd &	0.00 &	-0.04 &	-0.05 &	0.03 &	-0.02 &	0.01 \\
    Fe-Pd-Fe &	0.60 &	-0.21 &	0.39 &	0.59 &	-0.15 &	0.44 \\
    Fe-Pd-Pd &	0.36 &	-0.23 &	0.12 &	0.35 &	-0.17 &	0.18 \\
    Pd-Fe-Fe &	-0.43 &	0.24 &	-0.20 &	-0.47 &	0.32 &	-0.15 \\
    Pd-Fe-Pd &	-0.61 &	0.24 &	-0.37 &	-0.59 &	0.32 &	-0.28 \\
    Pd-Pd-Fe &	0.22 &	0.15 &	0.37 &	0.21 &	0.24 &	0.45 \\
    Pd-Pd-Pd &	0.13 &	0.11 &	0.24 &	0.12 &	0.21 &	0.33 \\
    \bottomrule
    \end{tabular}
\end{table}